# Multiple Criteria Clustering of Mobile Agents in WSN

Yashpal Singh[1] ,Kamal Deep[2] and S Niranjan[3]

[1]Research Scholar, Mewar University, Rajasthan, India
yashpalsingh009@gmail.com
[2]Assistant Professor, Department of Electronics and Communication ,JIET Jind,Haryana,india
erkdjangra@gmail.com
[3]Professor, Department of Computer Science & Engineering ,PDM College of Engineering Bahadurgarh ,Jhajjar,Haryana,India
niranjan.hig41@gmail.co

## ABSTRACT

*In Wireless sensor networks data aggregation with hundreds and thousands of sensor nodes is very complex task. Recently, mobile agents have been proposed for efficient data dissemination in sensor networks. In the traditional client/server based computing architecture, data is collected from multiple sources and forwarded to destination for further processing. It requires high bandwidth, whereas in the mobile agent is a task specific executable code traverses to the relevant source for gathering data. It reduces communication overhead, reduce cost, low bandwidth. Agents have capability to perform task for multiple applications. It will send only useful information to destination node. The problem is to group similar mobile agents into a number of clusters such that each cluster has similarity in responding to a group of nodes. By clustering intelligent mobile agents, it is possible to reduce the cost of time for each individual agent, decrease the demand imposed on network for a set of required tasks, decrease total number of visits. This paper, we present the problem of Multiple Criteria Clustering of Mobile Agents (MCCMA) where the decision is to cluster mobile agents such that a group of similar intelligent mobile agents will visit a group of similar sensor nodes.*

## 1. INTRODUCTION

Wireless Sensor Network (WSN) has received much attention in the research field in last few years. Sensors are expected to be inexpensive and can be deployed in a large scale over the network of operation. A fundamental concept of WSN is that all sensor nodes are distributed in specific area and them remains fixed in their respective position. Sink sensor node gather information about environment, compute it and then send it to clients. Energy-efficient data delivery and security is crucial as sensor nodes operate with limited battery power. Due to their flexibility and cost effectiveness, wireless sensor networks (WSNs) have been used for numerous applications including environmental monitoring, facility monitoring, and military surveillance for tasks such as target detection[17]. Although there has been an extensive research work done toward energy efficiency of WSN. In the traditional client/server based computing model, information exchange takes place amongst sensor nodes acting as source and sink node. In case a linked bandwidth of WSN is low and cannot meet network traffic and require consumption of more energy. A lot of research has been done on power utilization in WSN. Various researchers have identified different routing algorithms, scheduling algorithms, load balancing techniques, clustering schemes etc in order to improve life span of WSN. There also exist some problems in WSN:

- Path loss due to low bandwidth
- Loss of fusion accuracy
- More energy consumption

The data collectiona approachin WSN can be classified into multi path approaches, query propagation approaches, and mobile agent approaches. Multipath approaches achieve a high degree of reliability by making use of multiplepaths to send information from every sensor node to the collection point[15][16].

## 1.1 Mobile Agent concept

To meet the above challenges the concept of mobile agents has been proposed. To solve the problem of the overwhelming data traffic, bandwidth, Hairong et al, [1] proposed a mobile agent based distributed sensor network (MADSN) for scalable and energy-efficient data aggregation. Mobile Agents can be used in mobile computing environment for network control and management The Mobile Agent (MA) is a special kind of software which visits the network either periodically or on demand and performs data processing autonomously while migrating from node to node. There is very important property of mobile agents is their ability to autonomously move from one device to another (mobility). It is proved that mobile agent implementation can save up to 90 percents of data transfer time due to avoiding the raw data transfer, Feiyi Wang et al. [5]. Mobile agents are a distributed computing paradigm based on code mobility for high effectiveness and efficient in IP-based highly dynamic distributed environments A.R. Silva [2]. The two attributes SinkID and MA_SeqNum are used to identify an MA packet. Whenever the Sink sends a new MA packet, it increments the MA_SeqNum. The SrcList list
specifies the node itinerary LCF(Local Closest First ) to be visited by the MA. NextSrc specifically determines the sequence of node identifiers that must be visited by the MA[7][8]. All mobile agents have the features: timer, navigation algorithm, network information, and above all stability of the network and load traffic of network.

In traditional network domain, data is collected by the individual sensor and is transferred to sink node for further processing known as data fusion as in figure 1.1. So in this type of network a large amount of data travels between the network elements. By using mobile agent data only on demand need to be sent to any sink node as in figure 1.2. Mobile agent is proposed to perform the following functions: (1) eliminating data redundancy among sensors by local processing at node level in the data context; (2) eliminating spatial redundancy among closely-located sensors by data aggregation at the task level; (3) reducing communication overhead by concatenating data at the combined task level.

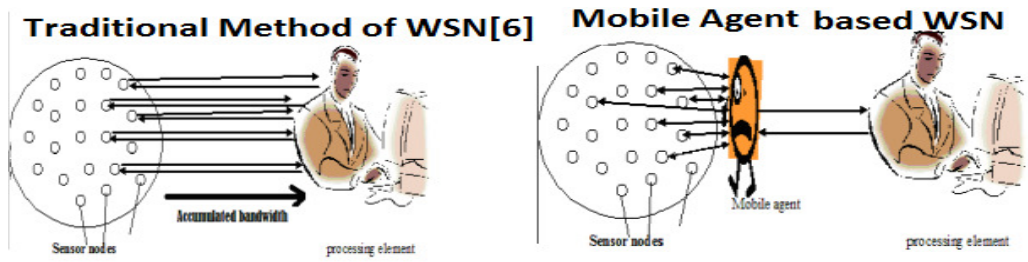

Fig. 1.1

Fig. 1.2

## 1.2 Work processing of Mobile Agents

The mobile agent to begin with when arrives at a node for the first time, stores its code at that node for future visit without carrying its code and sending the result to the sink node. Mobile agent also copies its processing code into the memory of each node in the first round. Once the whole task is completed, all those nodes discard the processing code. The agent performs the filtering function when it acquires the data. For example to estimate temperature, if node 1, node2 and node3 are neighbouring node, they must have similar temperature. We suppose that node1 estimate the temperature of thirty degree, node3 estimate the temperature of thirty two degree. node2 is located between node1 and node3. node2 must have the temperature either from thirty to thirty two or similar degree. But, if node2 estimates the temperature over forty or less than ten degree, Sensing data of nodes can be fault data. In this way agents can judge the data When the network connection is not available the mobile agent can save its status on secondary memory and later migrate to the home network when the connection is available. Software agents need not always travel across a network to communicate with information sources, or other agents. They work on message passing systems for agents.

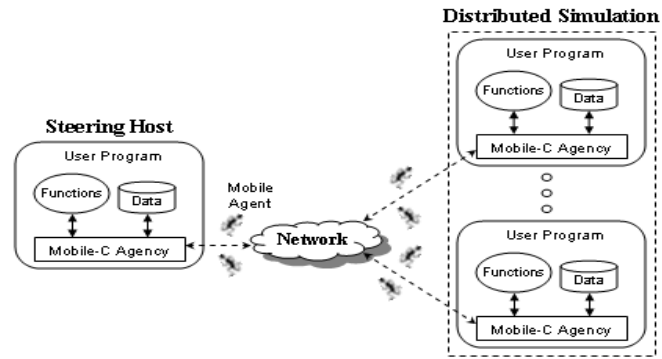

Figure 1.3 Multiple Criteria Clustering of Mobile Agents [9]

## 1.3 clustering of Mobile Agents

Cluster analysis is concerned with the grouping of alternatives into homogeneous clusters (groups) based on certain features. Three well-known clustering strategies are hierarchical clustering (Pandit, Srivastava, and Sharma 2001 ; Dias, Costa, and Climaco 1995), graph-theoretic methods (Matula 1977) and conceptual clustering (Michalski and Stepp 1982, 1983). Conventional clustering approaches involve two main factors: 1) distance (or dissimilarity) measurement, and 2) cluster centres. Head node has id information of nodes in its cluster. When the head node generates agents, the agent can have the id information of nodes that should pass. And the agent migrates to the nodes in pass route. And the agent also contains the information of neighboring nodes that are not included in the route. If node in route has the problems, the agent migrates to neighbouring node through the neighbouring information. By clustering mobile agents, cost of time for each agent to travel to its required routers, can be reduced. Routers have certain capabilities and can process certain requests of each intelligent agent.We propose in our present work a new application to clustering. We discuss how clustering methods can be applied and used for MCCMA communication network problems. Intelligent agents are grouped into cluster in such a way that each cluster has similarity in responding to a group of routers. In this way it helps to reduce demand imposed upon a network for a set of required task to be performed.

## 2. Related Work

Since sensor network works with limited energy,there has been an extensive research work toward energy efficiency of sensor. Liang Zhao [11] proposed a Medium-contention based Energy-efficient Distributed Clustering (MEDIC) scheme, through which sensors self-organize themselves into energy-efficient clusters by bidding for cluster headship. Sudhir [12] focuses on use of classification techniques using neural network to reduce the data traffic from the node and thereby reduce energy consumption. Low Energy Adaptive Clustering Hierarchy (LEACH) algorithm is applied to decompose the network into clusters, with one head for each. The main purpose of this technique is to collect the data by a mobile agent and to send them together to minimize the transmission . A number of papers have proposed algorithms for data Compression/Decompression (C/D) to reduce the amount of data transmitted by the sensors. The MA is a special kind of software that propagates over the network either periodically or on demand (when required by the applications). It performs data processing autonomously while migrating from node to node. Q. Wu et. al. [13]. In [14] *Costas Tsatsoulis et al* says that instead of one centralized and usually very large system that assumes the complete control and intelligence of the network, a number of smaller systems, or agents, can be used to help manage the network in a cooperative manner .This has motivated the multi agent systems (MAS) in telecommunication networks. The use of MAs in computer networks has certain advantages and disadvantages like code caching, safety and security, depending on the particular scenario. A lot of research has been done on power utilization in WSN various researchers have identified different clustering methods, routing algorithms, scheduling algorithms and load balancing algorithms.

In our proposed work, it is assumed that sensor network is divided into clusters. In traditional client/server approach, data is transmitted directly to sink node which will reduce the life span of sensor network. To overcome this approach clustering approach is used, in which sensors are clustered using clustering algorithm. Sensor nodes send all sensed data to sink node whether all data is necessary or not. This approach requires a lot of bandwidth and increase network traffic.

Mobile Agent is programmes which move sensor to sensor to gather information and transfer only specified information to sink node. This approach requires low bandwidth and reduce network traffic. Cost effective migration path is defined by the gateway as mobile agents visit the WSN. In this paper , Mobile Agents are divided into clusters based on multiple criteria. Then an incidence matrix is created and then agents are clustered using some rules defined below. The clustering of multiple criteria alternatives can also bring the following benefits:[9]

1) It decreases the set of alternatives - since the Decision Maker may be interested in only those alternatives with similar kinds of features and discard other alternatives.
2) It decreases the number of criteria - when evaluating alternatives of one group, the Decision Maker does not need to consider all criteria since one or more criteria values of the alternatives in the same group are equal or very close.
3) It provides a basis for more in-depth evaluation of alternatives - once one set of clustered alternatives are selected then this set can be explored in more depth for analysis, selection, and implementation purposes.
4) It may provide a basis for analyzing multiple criteria problems. Each decision maker may cluster alternatives differently, and hence, clustering of alternatives may provide a basis for negotiation.
5) In case of selection of a group of alternatives, each decision maker can be in charge of one clustered group;
Hence the designated decision maker selects the best solution from each clustered group *Miettnen and Salminen (1999) or Malakooti and Raman (2000).*

## 3. Clustering for Mobile Agents

There are two hierarchical levels of agents: load management agents and parent agents. The load management agent visits every node in the network efficiently using Dijkstra's shortest path algorithm and it collects the necessary information to determine the optimal routes from all other nodes to that particular node. The parent agents control the next management level. They travel around the network and launch load agents where network management is needed. In this paper we use load management agents for collecting needed information form each and every node.

Each router has certain capabilities to process certain requests or needs of each intelligent agent**.** The problem is to group the intelligent mobile agents into a number of clusters such that each cluster has similarity in meeting router demands. Suppose that an intelligent mobile agent must visit a certain number of routers in order to complete the task that it is assigned to. If required, additional agents can be assigned of identical capabilities. Such acquiring will increase cost .Hence each agent can be duplicated. In a simpler way, the procedure is to find cluster center Cr and then cluster all alternative objects into a group according some rules. First of all, we define similarity measurement d(X,C) as the generalized Euclidean distance between objects Xi and center Cr , r=1,2,3,4……..n[7]

$$d\,(X_i, Cr) = \sqrt{k_1(x_1 - c_{11})^2 + k_2(x_2 - c_{12})^2 + \dots + k_m(x_m - c_{1m})^2}$$

Where k1, k2……..$k_m$ are the generalized Euclidean distance, the coefficients k = {$k_i$:  i = 1, … m} can be  used  to represent indirectly the important  index of each attribute of the alternatives. An objective X belongs to cluster r if and only if d(X, Cr) < d(X, Ct) | t = 1, 2... R and t < r.

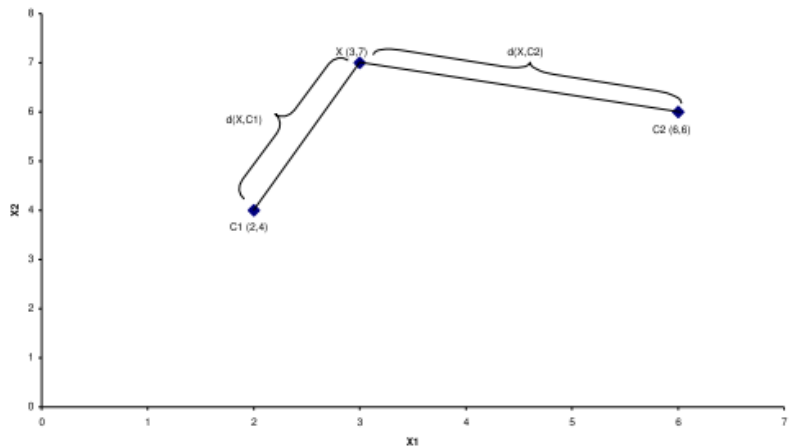

Figure 3.1 Example of Cluster Memberships

**Clustering Algorithm steps:**

Step 1:  Consider unclassified wireless sensor network area of (X*Y) meter having N number of routers and M number of agents.

Step 2:  Sink node dispatch mobile agents with requirement of specific task to wsn. Mobile agent migrate from first to last sensor node to perform specified task.

Step 3:  Route table of agents is created corresponding to their routers.

Step 4: A matrix {$A_{ij}$} called router-agent incidence matrix is created which has i rows of routers and j columns of agents. An element $a_{ij}$ of matrix {$A_{ij}$} is 1 if agent j require operation to be perform on router i, otherwise $a_{ij}$ is 0.

Step 5: Based on route table and incidence matrix of router-agent family, agents are divided into clusters. Agents perform operations on some specific routers are put into clusters.

Step 6: The few entries outside the diagonal blocks represent operations to be performed out-side the assigned group router cells. These elements are called exceptional elements.

We consider an example and divide into clusters

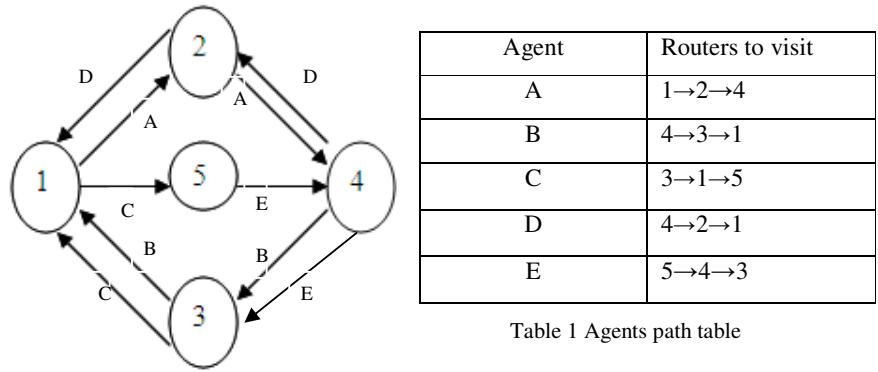

| Agent | Routers to visit |
|-------|------------------|
| A | 1→2→4 |
| B | 4→3→1 |
| C | 3→1→5 |
| D | 4→2→1 |
| E | 5→4→3 |

Table 1 Agents path table

Figure 3.2 Path of routers

| Cluster no. | Routers cells | Agents families |
|-------------|---------------|-----------------|
| Cluster 1 | 1,5,3 | B,C,E |
| Cluster 2 | 2,4 | A,D,E |

Table 2 (Agent family and visited Routers table)

Here Table 1 is divide the agents into two clusters. That is, the primary cluster for handling B,C,E agents corresponding router are 1,5,3 and for A,D,E agents corresponding routers are 2 and 4. Agents B,C,E perform some operation on router 1,5,3 and similarly A,D,E on router 2,4. Here in above example agent E used in both clusters so if we can afford the cost of more agents then we will acquire additional agents, the duplicated agents.

| Agents \ Routers | A | B | C | D | E |
|------------------|---|---|---|---|---|
| 1 | 1 | 1 | 1 | 1 | 0 |
| 2 | 1 | 0 | 0 | 1 | 0 |
| 3 | 0 | 1 | 1 | 0 | 1 |
| 4 | 1 | 0 | 0 | 1 | 1 |
| 5 | 0 | 0 | 1 | 0 | 1 |

Table 3 Incidence matrix of router-agent

The router-agent cell formation is a strategy to group routers into cells and agents into families. Families of agents can then be completely processed in their corresponding group router cells [10]. The processing of agents on routers can be represented in the form of a matrix called router-agent incidence matrix which has i rows of routers and j columns representing the agents.

## 4. Experimental work

The proposed work is shown in form of data flow diagram as follows:

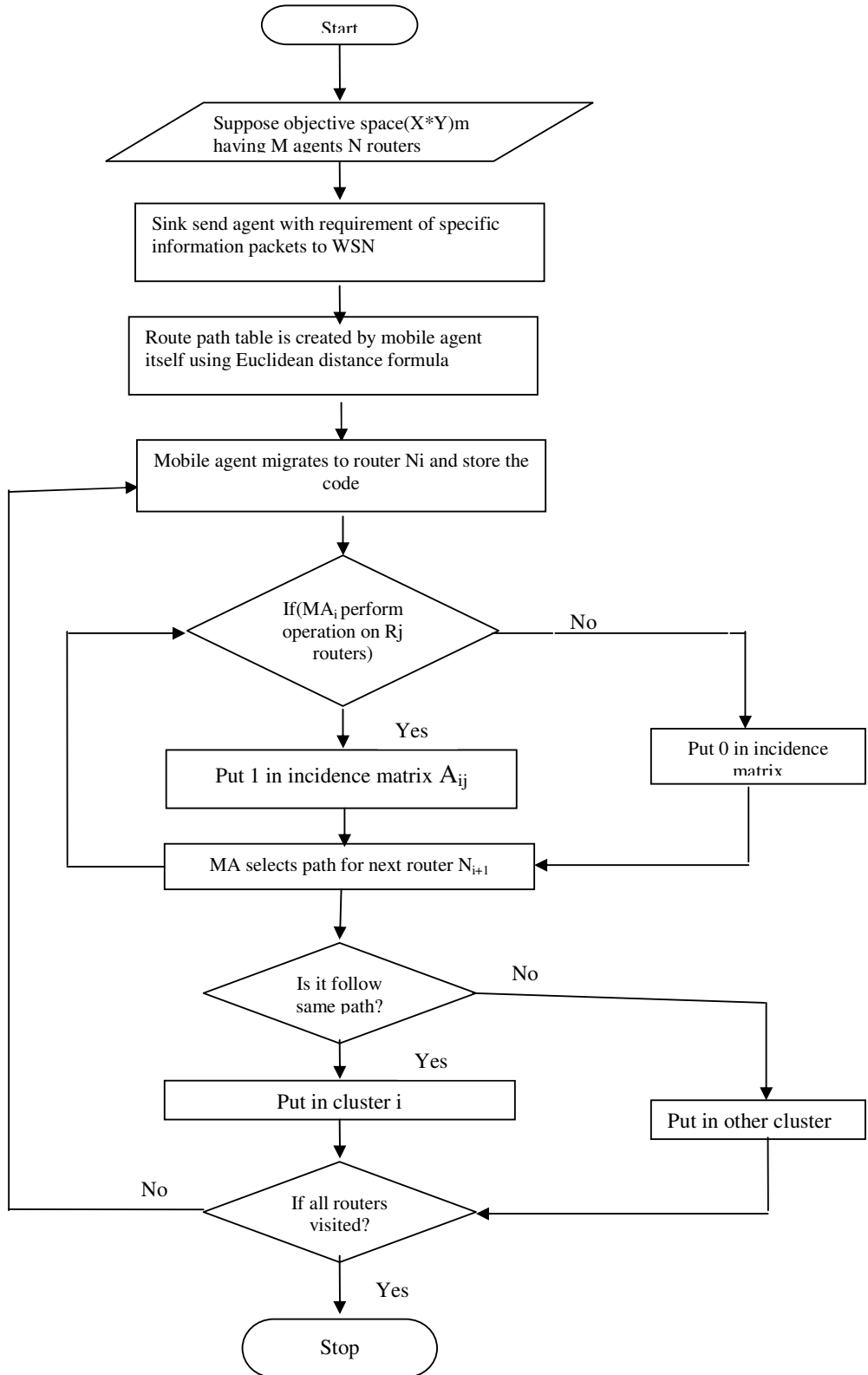

Now we take an example of 15*21 incidence matrix having 15routers and 21 agents grouping agents into families and routers into cells. Then result show block diagonal matrix which is clusters of agents based on some criteria stated above.

| Agent / Routers | 1 | 2 | 3 | 4 | 5 | 6 | 7 | 8 | 9 | 10 | 11 | 12 | 13 | 14 | 15 | 16 | 17 | 18 | 19 | 20 | 21 |
|---|---|---|---|---|---|---|---|---|---|---|---|---|---|---|---|---|---|---|---|---|---|
| 1 |  | 1 |  |  |  | 1 | 1 |  |  |  | 1 |  |  |  |  | 1 | 1 |  | 1 | 1 |  |
| 2 |  |  |  |  | 1 |  |  |  | 1 | 1 |  |  | 1 |  | 1 |  |  | 1 |  |  |  |
| 3 | 1 |  |  | 1 |  |  | 1 |  |  |  |  |  | 1 |  |  |  |  | 1 |  |  |  |
| 4 |  |  | 1 |  |  | 1 |  |  |  |  |  |  | 1 |  |  |  |  | 1 |  |  |  |
| 5 |  |  |  |  |  |  |  | 1 | 1 | 1 |  |  | 1 |  | 1 |  |  |  | 1 |  |  |
| 6 | 1 |  |  |  |  |  | 1 |  |  |  | 1 | 1 |  |  |  |  |  |  |  | 1 |  |
| 7 |  | 1 |  |  |  |  | 1 |  |  |  |  | 1 |  |  |  |  |  |  |  | 1 |  |
| 8 | 1 |  | 1 | 1 |  | 1 |  |  |  |  |  |  |  |  |  |  |  | 1 |  |  |  |
| 9 | 1 |  |  |  | 1 | 1 |  |  |  |  |  |  | 1 |  |  |  |  | 1 |  |  |  |
| 10 |  |  |  |  |  |  |  | 1 | 1 | 1 |  |  |  |  | 1 |  |  |  |  |  |  |
| 11 |  | 1 |  |  |  |  | 1 |  |  |  | 1 | 1 |  |  |  |  |  |  |  | 1 |  |
| 12 |  |  |  |  |  |  |  |  |  |  |  |  |  |  |  | 1 | 1 |  | 1 |  |  |
| 13 |  |  |  |  | 1 |  |  | 1 | 1 | 1 |  |  | 1 |  | 1 |  |  |  |  |  |  |
| 14 |  |  | 1 |  |  |  |  |  |  |  |  |  |  |  |  |  | 1 |  | 1 | 1 | 1 |
| 15 |  |  | 1 |  |  |  | 1 |  |  |  |  |  |  |  |  | 1 | 1 |  | 1 |  | 1 |

Table 5 Router-Agent initially incidence matrix

Now we divide above incidence matrix of agents and router according agent families and router cells as follows:

**Routers**   **Agents** ⟶

| Router | 2 | 7 | 11 | 12 | 20 | 1 | 3 | 4 | 6 | 14 | 18 | 5 | 8 | 9 | 10 | 13 | 15 | 16 | 17 | 19 | 21 |
|---|---|---|---|---|---|---|---|---|---|---|---|---|---|---|---|---|---|---|---|---|---|
| 1 | 1 | 1 | 1 | 0 | 1 |  |  |  |  |  |  |  |  |  |  |  |  |  |  | 1 |  |
| 6 | 0 | 1 | 1 | 1 | 1 |  |  |  |  |  |  |  |  |  |  |  |  |  |  |  |  |
| 7 | 1 | 1 | 0 | 1 | 1 |  |  |  |  |  |  |  |  |  |  |  |  |  |  |  |  |
| 11 | 1 | 1 | 1 | 1 | 1 |  |  |  |  |  |  |  |  |  |  |  |  |  |  |  |  |
| 3 |  |  |  |  |  | 1 | 0 | 1 | 0 | 1 | 1 |  |  |  |  |  |  |  |  |  |  |
| 4 |  |  |  |  |  | 0 | 1 | 0 | 1 | 1 | 1 |  |  |  |  |  |  |  |  |  |  |
|  |  |  |  |  |  | 1 | 1 | 1 | 1 | 0 | 1 |  |  |  |  |  |  |  |  |  |  |
| 9 |  |  |  |  |  | 1 | 0 | 0 | 1 | 1 | 1 |  |  |  |  |  |  |  |  |  |  |
| 2 |  |  |  |  |  |  |  |  |  |  |  | 1 | 0 | 1 | 1 | 1 | 1 |  |  |  |  |
| 5 |  |  |  |  |  |  |  |  |  |  |  | 0 | 1 | 1 | 1 | 1 | 1 |  | 1 |  |  |
| 10 |  |  |  |  |  |  |  |  |  |  |  | 1 | 1 | 1 | 1 | 0 | 1 |  |  |  |  |
| 13 |  |  |  |  |  |  |  |  |  |  |  | 1 | 1 | 1 | 1 | 1 | 1 |  |  |  |  |
| 12 |  |  |  |  |  |  |  |  |  |  |  |  |  |  |  |  |  | 1 | 1 | 1 | 0 |
| 14 |  |  |  | 1 |  |  |  |  |  |  |  |  |  |  |  |  |  | 0 | 1 | 1 | 1 |
| 15 |  | 1 |  |  |  | 1 |  |  |  |  |  |  |  |  |  |  |  | 1 | 1 | 1 | 1 |

Table 7 Clustered Matrix

| Cluster no. | Routers cells | Agents families |
|:---:|:---:|:---:|
| 1 | 1,6,7,11 | 2,7,11,12,20 |
| 2 | 2,5,10,13 | 5,8,9,10,13,15 |
| 3 | 3,4,8,9 | 1,3,4,6,14,18 |
| 4 | 12,14,15 | 16,17,19,21 |

Table 6(Agent family and visited Routers table)

The resultant matrix after grouping (according above table) has four distinct router-agent cells. Table 6 shows that agent family 1, consisting of agents 2, 7, 11, 12 and 20 can be processed in group router cell 1,which contains routers 1, 6, and 7 and 11 in similar ways family 2 ,consisting agents 1,3,4,6,14,18 process on routers 3,4,9 and in same way family3, 4. Here in table it is show that agents are clustered corresponding to their routers. The few entries outside the diagonal blocks represent operations to be performed outside the assigned group router cells. These elements are called exceptional elements. The corresponding router is called a bottleneck router, and the corresponding agent is called an exceptional agent.

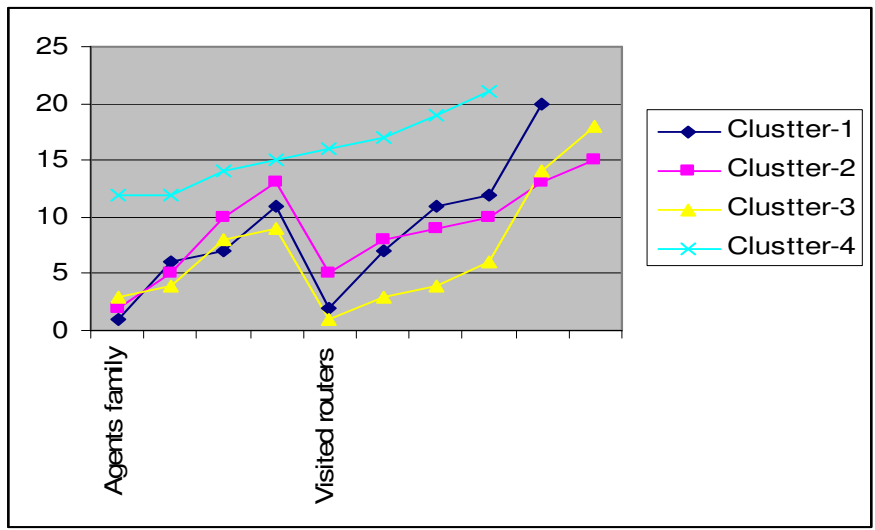

Figure 4.1 Clustering of router-Agent family

## 5. Conclusion

 In this paper, we discuss how clusters can be selected and how different alternatives could be clustered into independent clustered groups. Here our experimented result is shown in form of graph of router-agent family. Incidence matrix is divided into four clusters. In this paper, some fundamental definitions and algorithms are formulated for clustering intelligent mobile agents. We develop a method for Multiple Criteria Clustering of Mobile Agents (MCCMA) for wireless sensor network systems. Intelligent mobile agents were introduced as  agents those are sent to routers to carry out a set of required tasks. When clustering intelligent mobile agents with similar functional capabilities, the purpose is to minimize cost or travel time of each agent visiting its required routers. It will decrease network load for a set of required tasks to be performed. We demonstrate how intelligent agents are clustered into agent family and router cells by making a number of comparisons. The future development of this work includes clustering physical objects presented graphically to the decision maker; applying the selection of the most preferred alternative for each cluster, and considering the problem of multiple decision makers.

## Authors

**Mr. Yashpal Ssingh** obtained his MCA(2006) from G.J.U. University, M.Tech(CSE-2009) C.D.L.U University and Ph.D (CSE) Pursuing From Mewar University ,Rajasthan ,India. He has attended various national, International seminars, conferences and presented research papers on Artificial Intelligence ,Mobile Agent and Multi-Agent Technology.

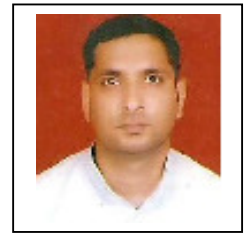

**Niranjan S** was born on 4[th] April 1955 in India. Graduated from University College of Engineering, Sambalpur University in 1978 in Electronics & Telecommunication Engineering, worked as asst Engineer under State Govt and Ex. Engineer under State Electricity Board working for Power line Carrier Communication Systems, Protective relaying, telemetry and LFC Systems. Master's Degree from IIT Kharagpur in 1987 in Computer Engineering. Worked in the areas of Parallel Processing. Performance characterization of parallel Programs under variable and uniprocessor environments, worked towards the development of a machine and hand printed Oriya Character Recognition System. Ph.D . from Utkal University. Area of work is Study and issues of mobile Computing Algorithms. He has also worked in the areas of Load Frequency Control in deregulated Scenarios and study of various non linear models of interconnected power Systems including GRC and other stochastic conditions. His current research areas include ECC based cryptographic applications, mobile ad-hoc sensor networks.

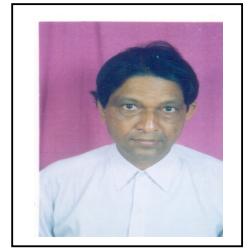

**Mr. Kamal Deep** obtained his BTech (2007) from KUK. University, M.Tech(-2011) M.D.U University and Ph.D (ECE) Pursuing From Mewar University ,Rajasthan ,India. He has attended various national, International seminars, conferences and presented research papers on WSN, Mobile Agent Technology. And working as a Assistant Professor in JIET,Jind ,India.

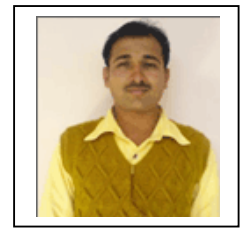